\documentstyle[sprocl,epsf]{article}

 \catcode`@=11
\def\@cite#1#2{\unskip\nobreak\relax
    \def\@tempa{$\m@th^{\hbox{\the\scriptfont0 #1}}$}%
    \futurelet\@tempc\@citexx}
\def\@citexx{\ifx.\@tempc\let\@tempd=\@citepunct\else
    \ifx,\@tempc\let\@tempd=\@citepunct\else
    \ifx;\@tempc\let\@tempd=\@citepunct\else
    \ifx:\@tempc\let\@tempd=\@citepunct\else
\let\@tempd=\@tempa\fi\fi\fi\fi\@tempd}
\def\@citepunct{\@tempc\edef\@sf{\spacefactor=\the\spacefactor\relax}\@tempa
    \@sf\@gobble}

\def\citenum#1{{\def\@cite##1##2{##1}\cite{#1}}}
\catcode`@=12

\def\lsim{\mathrel{\raise.2ex\hbox{$<$}\hskip-.8em\lower.9ex\hbox{$\sim$}}}
\def\gsim{\mathrel{\raise.2ex\hbox{$>$}\hskip-.8em\lower.9ex\hbox{$\sim$}}}

\begin{document}
\thispagestyle{empty}

\renewcommand{\thefootnote}{\fnsymbol{footnote}}

\vglue-.5in

\font\fortssbx=cmssbx10 scaled \magstep1
\hbox to \hsize{
\hbox{\fortssbx University of Wisconsin - Madison}
\hfill$\vtop{\hbox{\bf MADPH-99-1111}
            \hbox{April 1999}}$ }
\bigskip

\title{\uppercase{High Energy Neutrino Astronomy: WIN 99}}

\author{\unskip\medskip\uppercase{Francis Halzen}}

\address{Department of Physics, University of Wisconsin, Madison, WI 53706}

\maketitle\abstracts{
Although high energy neutrino astronomy is a multidisciplinary science, gamma ray bursts have become the theoretical focus since recent astronomical observations revealed their potential as cosmic particle accelerators. This spotlight is shared with investigations of the potential of high energy telescopes to observe oscillating atmospheric neutrinos. The Superkamiokande results have boosted atmospheric neutrinos from a calibration tool and a background for doing astronomy, to an opportunity to confirm the evidence for neutrino mass. Nevertheless, the highlights are mostly on the experimental front with the completion of the first-generation Baikal and AMANDA detectors. Neutrino signals from the Lake Baikal detector bode well for the flurry of activities in the Mediterranean. The completed AMANDA telescope announced first light, neutrinos actually, at this meeting.}

\section{Neutrino Astronomy: Multidisciplinary Science}

Using optical sensors buried in the deep clear ice or deployed in deep ocean and lake waters,  neutrino astronomers are attacking major problems in astronomy, astrophysics, cosmic ray physics and particle physics by commissioning a first generation of neutrino telescopes\cite{gaisser}. According to estimates covering a wide range of scientific objectives, a neutrino telescope with effective telescope area of 1 kilometer squared is required to address the most fundamental questions\cite{halzen}. Planning is already underway to instrument a cubic volume of ice, 1 kilometer on the side, as a neutrino detector. This infrastructure provides unique opportunities for yet more interdisciplinary science covering the geosciences and biology.

Among the many problems which high energy neutrino telescopes will address are the origin of cosmic rays, the engines which power active galaxies, the nature of gamma ray bursts (GRB), the search for the annihilation products of halo cold dark matter and, possibly, even the structure of the Earth's interior. In burst mode they scan the sky for galactic supernovae and, more speculatively, for the birth of supermassive black holes. Coincident experiments with Earth- and space-based gamma ray observatories, cosmic ray telescopes and gravitational wave detectors such as LIGO can be contemplated. With high-energy neutrino astrophysics we are poised to open a new window into space and back in time to the highest-energy processes in the Universe.

Neutrino telescopes can do particle physics. This is often illustrated by their capability to detect the annihilation into high energy neutrinos of neutralinos, the lightest supersymmetric particle which may constitute the cold dark matter. Also, with cosmological sources such as active galaxies and GRBs we will be observing $\nu_e$ and $\nu_\mu$ neutrinos over a baseline of $10^3$\,Megaparsecs. Above 1\,PeV these are absorbed by charged-current interactions in the Earth before reaching a detector at the opposite surface.  In contrast, the Earth never becomes opaque to $\nu_\tau$ since the $\tau$ produced in a charged-current $\nu_\tau$ interaction decays back into $\nu_\tau$ before losing significant energy\cite{saltzberg}. This penetration of tau neutrinos through the Earth above $10^2$\,TeV provides an experimental signature for neutrino oscillations. The appearance of a $\nu_{\tau}$ component in a pure $\nu_{e,\mu}$ beam would be signalled by a flat angular dependence of a source intensity at the highest neutrino energies.  Such a flat zenith angle dependence for the farthest sources is a signature for tau neutrino mixing with a sensitivity in $\Delta m^2$ as low as $10^{-17}$\,eV$^2$. With neutrino telescopes we will also search for ultrahigh-energy neutrino signatures from topological defects and magnetic monopoles; for properties of neutrinos such as mass, magnetic moment, and flavor-oscillations; and for clues to entirely new physical phenomena. The potential of neutrino ``telescopes" to do particle physics is evident.

\subsection{Cosmic Particle Accelerators: Gamma Ray Bursts Take Center Stage}

Recently, GRBs may have become the best motivated source for high energy neutrinos~\cite{waxman}.  Although neutrino emission may be less copious and less energetic than anticipated in some models of active galaxies, the predicted fluxes can be calculated in a relatively model-independent way. There is increasing observational support for a model where an initial event involving neutron stars, black holes or the collapse of highly magnetized rotating stars, deposits a solar mass of energy into a radius of order 100~km. Such a state is opaque to light. The observed gamma ray display is the result of a
relativistic shock which expands the original fireball by a factor $10^6$ in 1~second. Gamma rays arise by synchrotron radiation by relativistic electrons accelerated in the shock, possibly followed by inverse-Compton scattering.

It has been suggested~\cite{waxmanprime} that the same cataclysmic events produce the highest energy cosmic rays. This association is reinforced by more than the phenomenal
energy and luminosity. Both GRBs and the highest energy cosmic rays are produced in cosmological sources, {\it i.e.}, distributed throughout the Universe. Also, the average rate $\dot E \simeq 4\times10^{44}\rm~Mpc^{-3}~yr^{-1}$ at which energy is injected into the Universe as gamma rays from GRBs is similar to the rate at which energy must be injected in the highest energy cosmic rays in order to produce the observed cosmic ray flux beyond the ``ankle'' in the spectrum at $10^{18}$~eV.

\break

The association of cosmic rays with GRBs obviously requires that kinetic energy in the shock is converted into the acceleration of protons as well as electrons. It is assumed that the efficiency with which kinetic energy is converted to accelerated protons is comparable to that for electrons.  The production of high-energy neutrinos is inevitably a feature of the fireball model because the protons will photoproduce pions and, therefore, neutrinos in interactions with the gamma rays in the burst. We have a beam dump configuration where both the beam and target are constrained by observation: the beam by the observed cosmic ray spectrum and the photon target by astronomical measurements of the high energy photon flux.

From an observational point of view, the predicted flux can be summarized in terms of the main ingredients of the model:
\begin{equation}
N_\nu (km^{-2}year^{-1}) \simeq 25 \left[ f_\pi\over 20\% \right] \left[ \dot E\over 4\times10^{44}
{\rm\, Mpc^{-3} \, yr^{-1}} \right] \left[ E_\nu\over 700\rm\ TeV
\right]^{-1} \,,
\end{equation}
{\it i.e.},
we expect 25 events in a km$^2$ detector in one year. Here $f_{\pi}$, estimated to be 20\%, is the efficiency by which proton energy is converted into the production of pions and $\dot E$ is the total injection rate into GRBs averaged over volume and time. The energy of
the neutrinos is fixed by particle physics and is determined by the threshold for photoproduction of pions by the protons on the GRB photons in the shock. Note that GRBs produce a ``burst" spectrum of neutrinos. After folding the falling GRB energy spectrum with the increasing detection efficiency, a burst energy distribution results centered on an average energy of several hundred TeV.

Interestingly, this flux may be observable in operating first-generation detectors. The effective area for the detection of 100~TeV neutrinos can approach 0.1~km$^2$ as a result of the large size of the events. A $\nu_e$ of this energy initiates an electromagnetic shower which produces single photoelectron signals in ice over a radius of 250~m. The effective area for a $\nu_{\mu}$ is larger because the muon has a range of 10~km (water-equivalent) and produces single photoelectrons as far as 200~m from the track by catastrophic energy losses. Because these spectacular events arrive with the GRB time-stamp of order 1 second precision and with an unmistakable high-energy signature, background rejection is greatly simplified. Although, on average, we expect much less than one event per burst, a relatively near burst would produce multiple events in a single second. The considerable simplification of observing high energy neutrinos in the burst mode has inspired proposals for highly simplified dedicated detectors\cite{crawford}.

\break

The importance of making GRB observations cannot be overemphasized\cite{waxman}:
\begin{itemize}
\item The observations are a direct probe of the fireball model of GRBs.
\item They may unveil the source of the highest energy cosmic rays.
\item The zenith angle distribution of the GRB neutrinos may reveal the appearance of $\nu_{\tau}$ in what was a $\nu_e, \nu_{\mu}$ beam at its origin. The appearance experiment with a baseline of thousands of megaparsecs has a sensitivity to oscillations of $\Delta m^2$ as low as $10^{-17}$\,eV$^2$, as previously discussed.
\item The relative timing of photons and neutrinos over cosmological distances will allow unrivaled tests of special relativity.
\item The fact that photons and neutrinos should suffer the same time delay travelling through the gravitational field of our galaxy will lead to better tests of the weak equivalence principle.
\end{itemize}

In response to the evidence that atmospheric neutrinos oscillate\cite{superK}, workers in this field have been investigating the possibility of studying neutrino mass with the atmospheric neutrinos which, up to now, are used for calibration only. These studies may significantly reshape the architecture of some detectors. We will return to this topic later on.

\section{Large Natural Cherenkov Detectors}

The study of GRBs is one more example of a science mission that requires kilometer-scale neutrino detectors. This is not a real surprise. The probability to detect a PeV neutrino is roughly $10^{-3}$. This is easily computed from the requirement that, in order to be detected, the neutrino has to interact within a distance of the detector which is shorter than the range of the muon it produces\cite{gaisser}. At PeV energy the cosmic ray flux is of order 1 per m$^{2}$ per year and the probability to detect a neutrino of this energy is of order 10$^{-3}$. A neutrino flux equal to the cosmic ray flux will therefore yield only a few events per day in a kilometer squared detector. At EeV energy the situation is worse. With a cosmic ray rate of 1 per km$^2$ per year and a detection probability of 0.1, one can only detect several events per year in a kilometer squared detector provided the neutrino flux exceeds the proton flux by 2 orders of magnitude or more. For the neutrino flux generated by cosmic rays interacting with CMBR photons and for sources like active galaxies and topological defects\cite{schramm}, this is indeed the case. All above estimates are however conservative and the rates should be higher because absorption of protons in the source is expected, and the neutrinos escape the source with a flatter energy spectrum than the protons. In summary, at least where cosmic rays are part of the beam dump, their flux and the neutrino cross section and muon range define the size of a neutrino telescope. A telescope with kilometer squared effective area represents a neutrino detector of kilometer cubed volume.

The first generation of neutrino telescopes, launched by the bold decision of the DUMAND collaboration over 25 years ago to construct such an instrument, are designed to reach a large telescope area and detection volume for a neutrino threshold of order 10~GeV. This relatively low threshold permits calibration of the novel instrument on the known flux of atmospheric neutrinos.  The architecture is optimized for reconstructing the Cherenkov light front radiated by an up-going, neutrino-induced muon. Only up-going muons made by neutrinos reaching us through the Earth can be successfully detected. The Earth is used as a filter to screen the fatal background of cosmic ray muons. This makes neutrino detection possible over the lower hemisphere of the detector. Up-going muons must be identified in a background of down-going, cosmic ray muons which are more than $10^5$ times more frequent for a depth of 1$\sim$2 kilometers. The method is sketched in Fig.\,1.

\begin{figure}[h]
\centering
\hspace{0in}\epsfxsize=2.25in\epsffile{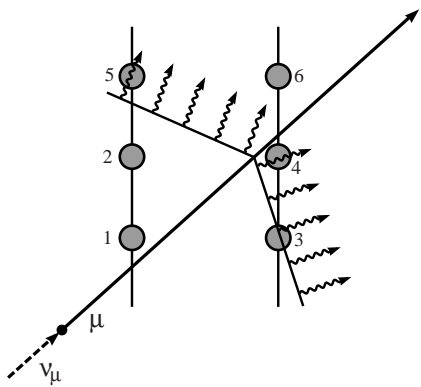}

\caption{The arrival times of the Cherenkov photons in 6 optical sensors determine the direction of the muon track.}
\end{figure}

The optical requirements of the detector medium are severe. A large absorption length is required because it determines the spacings of the optical sensors and, to a significant extent, the cost of the detector. A long scattering length is needed to preserve the geometry of the Cherenkov pattern. Nature has been kind and offered ice and water as adequate natural Cherencov media. Their optical properties are, in fact, complementary. Water and ice have similar attenuation length, with the role of scattering and absorption reversed; see Table~1. Optics seems, at present, to drive the evolution of ice and water detectors in predictable directions: towards very large telescope area in ice exploiting the large absorption length, and towards lower threshold and good muon track reconstruction in water exploiting the large scattering length.

\begin{table}[t]
\def\arraystretch{1.5}\tabcolsep=1em
\caption{Optical properties of South Pole ice at 1750\,m, Lake Baikal water at 1\,km, and the range of results from measurements in ocean water below 4\,km.}
\smallskip
\centering\leavevmode
\begin{tabular}{lccc}
\hline
& (1700 m)&&\\[-1.5ex]
$\lambda = 385$ nm\,$^*$& AMANDA& BAIKAL& OCEAN\\
\hline
attenuation& $\sim 30$ m\,$^{**}$& \llap{$\sim$}8 m& 25--30 m\,$^{***}$\\
absorption& 100 m~~ & 8 m& ---\\
scattering& 25 m& 150--300 m& ---\\[-2ex]
length&&&\\
\hline
\multicolumn{4}{l}{\llap{$^*$}\,peak PMT efficiency}\\[-1ex]
\multicolumn{4}{l}{\llap{$^{**}$}\,same for bluer wavelengths}\\[-1ex]
\multicolumn{4}{l}{\llap{$^{***}$}\,smaller for bluer wavelengths}
\end{tabular}
\end{table}

\section{Baikal, ANTARES, Nestor and NEMO: Northern Water.}

Whereas the science is compelling, the real challenge is to develop a reliable, expandable and affordable detector technology. With the termination of the pioneering DUMAND experiment, the efforts in water are, at present, spearheaded by the Baikal experiment\cite{domogatsky}. The Baikal Neutrino Telescope is deployed in Lake Baikal, Siberia, 3.6\,km from shore at a depth of 1.1\,km. An umbrella-like frame holds 8 strings, each instrumented with 24 pairs of 37-cm diameter {\it QUASAR} photomultiplier tubes (PMT). Two PMTs in a pair are switched in coincidence in order to suppress background from natural radioactivity and bioluminescence. Operating with 144 optical modules since April 1997, the {\it NT-200} detector has been completed in April 1998 with 192 optical modules (OM). The Baikal detector is well understood and the first atmospheric neutrinos have been identified. 

The Baikal site is competitive with deep oceans although the smaller absorption length of Cherenkov light in lake water requires a somewhat denser spacing of the OMs. This does however result in a lower threshold which is a definite advantage, for instance for oscillation measurements and WIMP searches. They have shown that their shallow depth of 1 kilometer does not represent a serious drawback. By far the most significant advantage is the site with a seasonal ice cover which allows reliable and inexpensive deployment and repair of detector elements.

With data taken with 96 OMs only, they have shown that atmospheric muons can be reconstructed with sufficient accuracy to identify atmospheric neutrinos; see Fig.\,2. The neutrino events are isolated from the cosmic ray muon background by imposing a restriction on the chi-square of the Cherenkov fit, and by requiring consistency between the reconstructed trajectory and the spatial locations of the OMs reporting signals. In order to guarantee a minimum lever arm for track fitting, they were forced to reject events with a projection of the most distant channels on the track smaller than 35 meters. This does, of course, result in a higher threshold.

\begin{figure}[h]
\centering\leavevmode
\epsfxsize=2.9in\epsffile{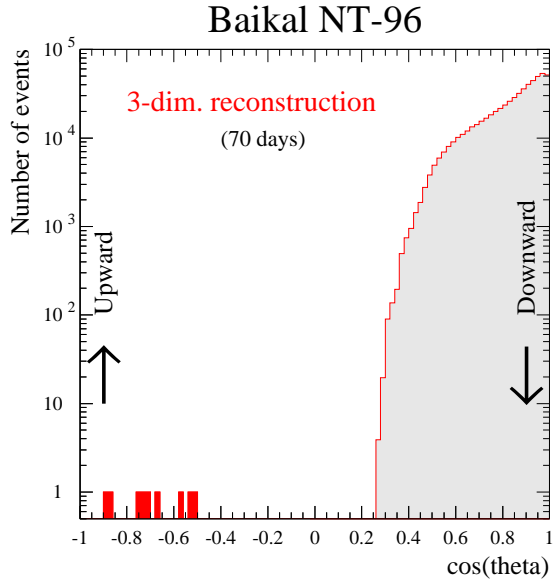}

\caption{Angular distribution of muon tracks in the Lake Baikal experiment after the cuts described in the text.}
\end{figure}

In the following years, {\it NT-200} will be operated as a neutrino telescope with an effective area between $10^3 \sim 5\times 10^3$\,m$^2$, depending on energy. Presumably too small to detect neutrinos from extraterrestrial sources, {\it NT-200} will serve as the prototype for a larger telescope. For instance, with 2000 OMs, a threshold of  $10 \sim 20$\,GeV and an effective area of $5\times10^4 \sim 10^5$\,m$^2$, an expanded Baikal telescope would fill the gap between present underground detectors and planned high threshold detectors of cubic kilometer size. Its key advantage would be low threshold.

The Baikal experiment represents a proof of concept for deep ocean projects. These have the advantage of larger depth and optically superior water. Their challenge is to find reliable and affordable solutions to a variety of technological challenges for deploying a deep underwater detector. Several groups are confronting the problem, both NESTOR and ANTARES are developing rather different detector concepts in the Mediterranean.

The NESTOR collaboration\cite{resvanis}, as part of a series of ongoing technology tests, is testing the umbrella structure which will hold the OMs. They have already deployed two aluminum ``floors", 34\,m in diameter, to a depth of 2600\,m. Mechanical robustness was demonstrated by towing the structure, submerged below 2000\,m, from shore to the site and back. These test should soon be repeated with fully instrumented floors. The actual detector will consist of a tower of 12 six-legged floors vertically separated by 30\,m. Each floor contains 14 OMs with four times the photocathode area of the commercial 8~inch photomultipliers used by AMANDA and ANTARES.

The detector concept is patterned along the Baikal design. The symmetric up/down orientation of the OMs will result in uniform angular acceptance and the relatively close spacings in a low threshold. NESTOR does have the advantage of a superb site, possibly the best in the Mediterranean. The detector can be deployed below 3.5\,km relatively close to shore. With the attenuation length peaking at 55\,m near 470\,nm the site is optically superior to that of all deep water sites investigated for neutrino astronomy.

The ANTARES collaboration\cite{feinstein} is investigating the suitability of a 2400\,m deep Mediterranean site off Toulon, France. The site is a trade-off between acceptable optical properties of the water and easy access to ocean technology. Their detector concept requires, for instance, ROV's for making underwater connections. First results on water quality are very encouraging with an attenuation length of 40\,m at 467\,nm and a scattering length exceeding 100\,m. Random noise exceeding 50\,khz per OM is eliminated by requiring coincidences between neighboring OMs, as is done in the Lake Baikal design. Unlike other water experiments, they will point all photomultipliers sideways or down in order to avoid the effects of biofouling. The problem is significant at the Toulon site, but only affects the upper pole region of the OM. Relatively weak intensity and long duration bioluminescence results in an acceptable deadtime of the detector. They have demonstrated their capability of deploying and retrieving a string.

With the study of atmospheric neutrino oscillations as a top priority, they plan to deploy in 2001-2003 10 strings instrumented over 400\,m with 100 OMs. After study of the underwater currents they decided that they can space the strings by 100\,m, and possibly by 60\,m. The large photocathode density of the array will allow the study of oscillations in the range $255 < L/E < 2550 \rm~km\,GeV^{-1}$ with neutrinos in the energy range $5 < E_{\nu} < 50$~GeV.

A new R\&D initiative based in Catania, Sicily has been mapping Mediterranean sites, studying mechanical structures and low power electronics. One must hope that with a successful pioneering neutrino detector of $10^{-3}\rm\, km^3$ in Lake Baikal, a forthcoming $10^{-2}\rm\, km^3$ detector near Toulon, the Mediterranean effort will converge on a $10^{-1}\rm\, km^3$ detector at the NESTOR site\cite{spiro}. For neutrino astronomy to become a viable science several of these, or other, projects will have to succeed besides AMANDA. Astronomy, whether in the optical or in any other wave-band, thrives on a diversity of complementary instruments, not on ``a single best instrument". When, for instance, the Soviet government tried out the latter method by creating a national large mirror project, it virtually annihilated the field.

\section{AMANDA: Southern Ice}

Construction of the first-generation AMANDA detector was completed in the austral summer 96--97. It consists of 300 optical modules deployed at a depth of 1500--2000~m; see Fig.\,3. Here the optical module consists of an 8~inch photomultiplier tube and nothing else. It is connected to the surface by a cable which transmits the HV as well as the anode current of a triggered photomultiplier. The instrumented volume and the effective telescope area of this instrument matches those of the ultimate DUMAND Octogon detector which, unfortunately, could not be completed.

\begin{figure}[t]
\centering\leavevmode
\hspace{0in}\epsfxsize=4in%
\epsffile{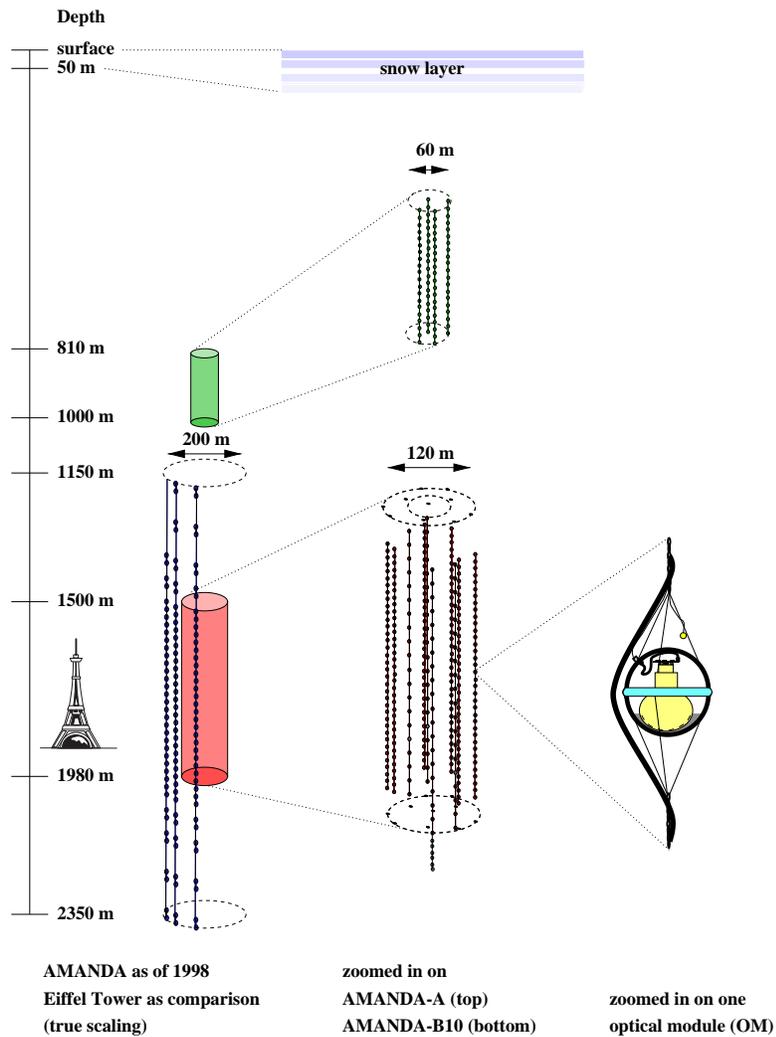}

\caption{The Antarctic Muon And Neutrino Detector Array (AMANDA).}
\end{figure}

As predicted from transparency measurements performed with strings near 1\,km depth\cite{science}, it was found that ice is bubble-free below 1400\,m. Its optical properties were a surprise, nevertheless the bubble-free ice turned out to be an adequate Cherenkov medium. We explain this next.

The AMANDA detector was antecedently proposed on the premise that inferior
properties of ice as a particle detector with respect to water could be compensated by additional optical modules. The technique was supposed to be a
factor $5 {\sim} 10$ more cost-effective and, therefore, competitive. The
design was based on then current information\cite{dublin}:
\begin{itemize}
\item
the absorption length at 370~nm, the wavelength where photomultipliers are
maximally efficient, had been measured to be 8~m;
\item
the scattering length was unknown;
\item
the AMANDA strategy was to use a large number of closely spaced OM's to overcome the short absorption length. Simulations indicated that muon tracks triggering 6 or more OM's were reconstructed with degree accuracy. Taking data with a simple majority trigger of 6 OM's or more at 100~Hz resulted in an average effective telescope area of $10^4$~m$^2$, somewhat smaller for atmospheric neutrinos and significantly larger for the high energy signals.
\end{itemize}

\noindent
The reality is that:
\begin{itemize}
\item
the absorption length is 100~m or more, depending on depth\cite{science};
\item
the scattering length is ${\sim} 25$~m (preliminary, this number represents an average value which may include the combined effects of deep ice and the refrozen ice disturbed by the hot water drilling);
\item
because of the large absorption length, OM spacings are similar, actually
larger, than those of proposed water detectors. Also, in a trigger 20 OM's report, not 6. Of these more than 6 photons are, on average, ``not scattered\rlap." A ``direct" photon is typically required to arrive within 25\,nsec of the time predicted by the Cherenkov fit. This allows for a small amount of scattering and includes the dispersion of the anode signals over the 2\,km cable. In the end, reconstruction is therefore as before, although additional information can be extracted from scattered photons by minimizing a
likelihood function which matches their measured and expected time delays.
\end{itemize}

The most striking demonstration of the quality of natural ice as a Cherenkov detector medium is the observation of atmospheric neutrino candidates with the partially deployed AMANDA detector which consisted of only eighty 8 inch photomultiplier tubes\cite{B4}. The up-going muons are separated from the down-going cosmic ray background once a sufficient number of direct photons and a minimum track length guarantee adequate reconstruction of the Cherenkov cone. For details, see Ref.~\citenum{B4}. The analysis methods were verified by reconstructing cosmic ray muon tracks registered in coincidence with a surface air shower array.

After completion of the AMANDA detector with 300 OMs, a similar analysis led to a first calibration of the instrument using the atmospheric neutrino beam. The separation of signal and background are shown in Fig.\,4 after requiring, sequentially, 5 direct photons, a minimum 100\,m track length, and 6 direct photons per event. The details are somewhat more complicated; see Ref.~\citenum{albrecht}. A neutrino event is shown in Fig.\,5. By requiring the long muon track the events are gold-plated, but the threshold high, roughly $E_{\nu} \geq 50$~GeV. This type of analysis will allow AMANDA to harvest of order 100 atmospheric neutrinos per year, adequate for calibration.

\begin{figure}[t]
\centering\leavevmode
\epsfxsize=2.7in\epsffile{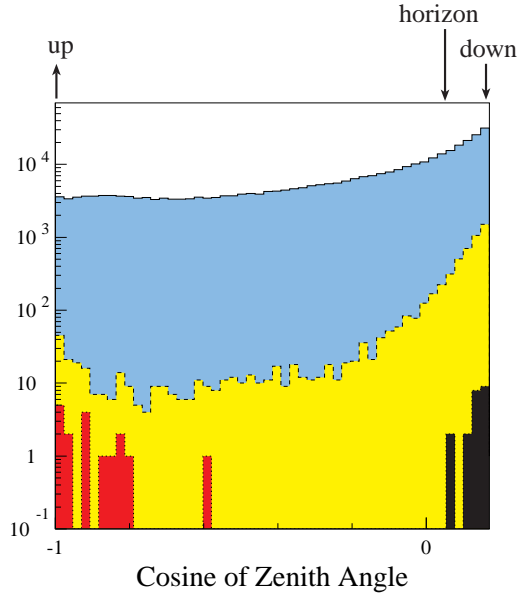}

\caption[]{Angular distribution of muon tracks in AMANDA at three levels of quality cuts. Roughly, $N_{\rm direct}\geq 5$, muon track longer than 100~m, and $N_{\rm direct} \geq 6$. Details in Ref.~\citenum{albrecht}.}
\end{figure}

While calibration continues, science started. Ongoing efforts cover a wide range of goals: indirect search for dark matter, active galaxies, gamma ray bursts and magnetic monopoles.

\looseness=-1
While water detectors exploit large scattering length to achieve sub-degree angular resolution, ice detectors can more readily achieve large telescope area because of the long absorption length of blue light. By instrumenting a cube of ice, 1~kilometer on the side, the planned IceCube detector will reach the effective telescope area of 1 kilometer squared which is, according to estimates covering a wide range of scientific objectives, required to address the most fundamental questions\cite{halzen}. A strawman detector with effective area in excess of 1\,km$^2$ consists of 4800\,OM's: 80 strings spaced by $\sim$\,100\,m, each instrumented with 60\,OM's spaced by 15\,m. IceCube will offer great advantages over AMANDA and AMANDA II\cite{albrecht} beyond its larger size: it will have a much higher efficiency to reconstruct tracks, map showers from electron- and tau-neutrinos (events where both the production and decay of a $\tau$ produced by a $\nu_{\tau}$ can be identified) and, most importantly, adequately measure neutrino energy.

\begin{figure}
\centering\leavevmode
\epsfxsize=4.7in\epsffile{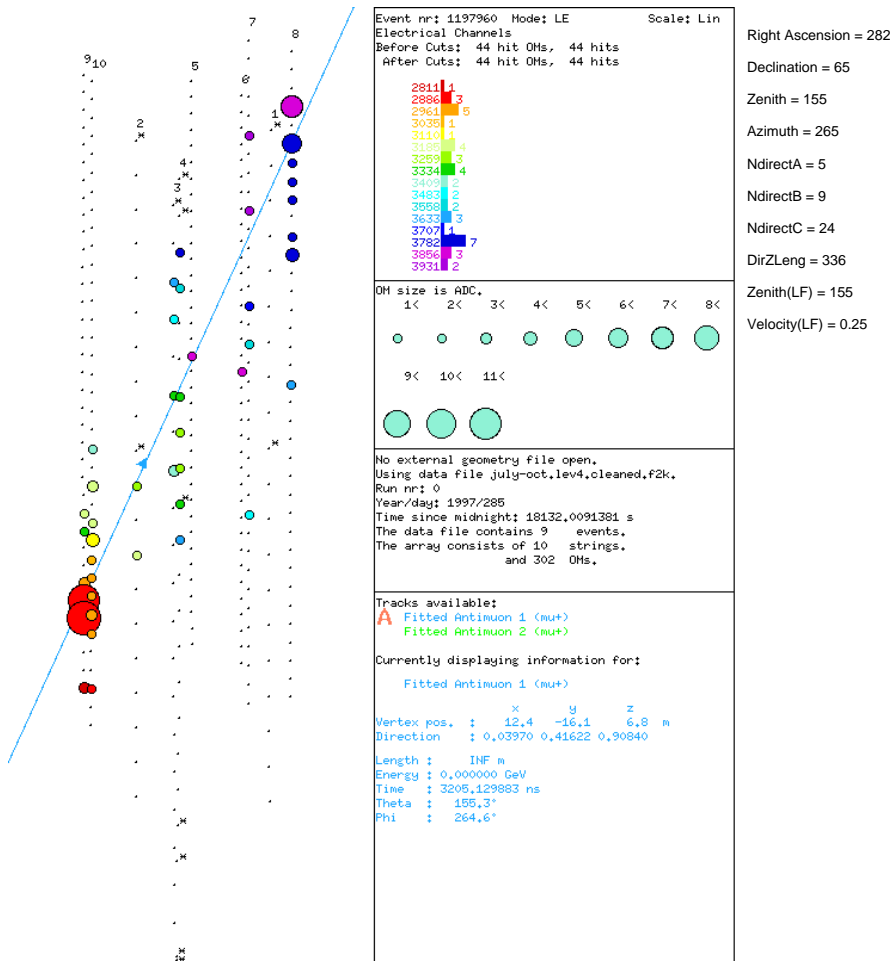}

\caption{Neutrino candidate in AMANDA. The shading (size) of the dots represents time (amplitude) of triggered photomultipliers. The reconstructed muon track moves upward over more than 300~m.}
\end{figure}

\newpage
\noindent{\bf Acknowledgments:} \ This work was supported in part by the University of Wisconsin Research Committee with funds granted by the Wisconsin Alumni Research Foundation, and in part by the U.S.\,Department of Energy under Grant No.\,DE-FG02-95ER40896.

\end{document}